# Studying the Impact of Power Capping on MapReduce-based, Data-intensive Mini-applications on Intel KNL and KNM Architectures

Extended Abstract

Joshua Hoke Davis
University of Delaware
jhdavis@udel.edu

Tao Gao (advisor)
National University of Defense Technology
gaotao@nudt.edu.cn

Sunita Chandrasekaran (advisor)
University of Delaware
schandra@udel.edu

Michela Taufer (advisor)
University of Tennessee
taufer@acm.org

## ABSTRACT

In this poster, we quantitatively measure the impacts of data movement on performance in MapReduce-based applications when executed on HPC systems. We leverage the PAPI 'powercap' component to identify ideal conditions for execution of our applications in terms of (1) dataset characteristics (i.e., unique words); (2) HPC system (i.e., KNL and KNM); and (3) implementation of the MapReduce programming model (i.e., with or without combiner optimizations). Results confirm the high energy and runtime costs of data movement, and the benefits of the combiner optimization on these costs.

## KEYWORDS

Performance studies of HPC subsystems, Energy efficiency, Workload characterization

## 1. Introduction

Data analytics and data-intensive workloads are gaining representation at peta- and exascale, and the MapReduce programming model is one model considered by the HPC community. It is a common belief that moving data is these workloads is power-intensive. There is the need to quantitatively measure the impacts of data movement on performance in MapReduce-based applications when executed on HPC systems. In this project, we study the impact of power capping on performance metrics such as runtime and energy consumption of a data-intensive application on top of a MapReduce over MPI framework. Our key questions are: 1) to identify the effects of power capping on the performance of data-intensive applications like MapReduce; and 2) to determine if there is an acceptable trade-off between power usage and performance for these applications.

## 2. MapReduce Programming Model and Mimir

We target data-intensive applications using the MapReduce (MR) programming model [1]. MR allows the user to provide a *map* and a *reduce* function, while handling the parallel job execution, communication, and data movement. MapReduce has three stages: the map processes a chunk of input data, converting the data to *<key, value>* (KV) pairs; the shuffle stage is an all-to-all communication which converts the KVs to *<key, <value1, value2, ...>>* (KMV) lists; the reduce takes KMVs and merges the values. Our MapReduce implementation is Mimir, a HPC fully in-memory MR over MPI [2].

## 3. Our Approach

We implement and use three data-intensive mini-apps extracted from the wordcount benchmark: (1) Map+Shuffle, which contains only the map and shuffle stages; (2) GroupByKey, which adds the reduce stage for a complete MapReduce execution; and (3) ReduceByKey, which locally combines KVs with matching keys immediately before shuffling (i.e., MR combiner).

We run our mini-apps on two manycore systems: the Intel Xeon Phi 7250 Knights Landing (KNL), and the Xeon Phi 7295 Knights Mill (KNM). The KNL machine features 68 cores, and a thermal design power (TDP) of 215 watts; the KNM has 72 cores at 320 watts TDP. We use the systems in FLAT mode, with all allocations happening in DRAM, and not in MCDRAM. We use PAPI, the Performance Application Programing Interface, to set power limits and collect energy consumption, using Intel RAPL (Running Average Power Limit) [3]. We conduct measurements with a 100 millisecond sample rate. We generate synthetic data (i.e., random words) which fully fit in the system memory, avoiding disk IO during our tests. Our datasets have two characteristics: a variable total number of words to count, and a variable number of unique words these are made up of. The number of unique words determines how the KMV are distributed across processes during shuffle. The number of unique words also impacts the the percentage of the dataset that can be combined before shuffling in ReduceByKey, referred to hereafter as combinability. We replicate each test three times, with the first of each run shown on graphs due to the modest variability across the three runs.

## 4. Results and Discussion

We measure power usage over time, total energy and runtime of our mini-apps. Figure 1 shows the performance metrics for the three mini-apps on KNM using a dataset of 4 billion total words and 72 unique words. We draw three pairs of curves for three power caps: 215, 140, and 120 watts. For each pair, higher curve is the processor power, and the lower is the DRAM power. Note that other measurements are in the poster.

Looking at the poster, we note that none of our tests exceed 160 watts during runtime (75 watts below system TDP). To study performance impacts, we intentionally impose caps of 140 and 120 watts, which are lower than the application's max power.

Between the KNL and KNM systems we observe the same pattern of power usage; however, tests on KNM have a lower runtime, especially at power limits that are below the normal power consumption of the benchmark. This is due KNM's higher number of cores.

The DRAM power forms only a small portion of total power use, only making up between 4% and 15% of the combined processor and DRAM power. Memory power consumption increases during the map and shuffle stages of GroupByKey (and in the Map+Shuffle benchmark), and during ReduceByKey tests when the combinability is low.

For Map+Shuffle and GroupByKey, when limiting power to 120 watts, we observe an increase in runtime ranging from 5% to 33%. Comparing Map+Shuffle and GroupByKey allows us to quantify the high runtime and energy costs of the reduce stage. The reduce stage adds 333%-355% more runtime and 335%-359% more energy to Map+Shuffle.

Comparing GroupByKey to ReduceByKey allows us to quantify the cost of moving data between processes. Combining KVs before shuffling in ReduceByKey substantially reduces the runtime of the application (up to 46% less runtime), and the energy consumption (up to 11,800 joules saved, or a 50% reduction). Most importantly, the combiner optimization introduces an implicit power limit without PAPI's intervention.

When zooming into the combiner performance for datasets with different numbers of unique words (and thus different combinabilities), the runtime and energy consumption vary around a 'sweet spot' region, a region of minimum runtime and energy. This 'sweet spot' region is around 4000 to 5000 unique words for our 4-billion-word dataset. For larger numbers of unique words, the memory and processor power increase and begin to oscillate. This oscillatory pattern is due to the lower effectiveness of the combiner on datasets with more diverse words, leading to to more KVs being moved during the Shuffle phase, and more frequent memory accesses. For smaller numbers of unique words, which create a highly homogeneous dataset, we encounter a latency associated with the inability to fill the MPI communication buffers.

## 5. Conclusion and Future Work

In this poster, we quantitatively measure the impact of data movement across cores on both runtime and energy consumption for a set of mini-apps on two manycore systems (KNL and KNM). Among our observations, we notice how the combiner optimization leads to up to a 46% reduction in runtime and a 50% reduction in energy consumption, without the need for a power cap. Our work-in-progress includes: (1) understanding how far our observations are from a general principle relating power cap and performance; (2) studying ways of reducing data movement other than the combiner used in this poster; and (3) understanding how the settings of the underlying MR can be tuned during runtime to extend the 'sweet spot' region.

Our artifact description is availible at https://drive.google.com/file/d/19XMWfGuolClKFnLN887LtPsffrGuVYUT/view?usp=sharing. The associated poster is at https://drive.google.com/file/d/1qDaDUq1pP7nzhHkdYKkz-gJYpHsBPhER/view?usp=sharing.

## ACKNOWLEDGEMENTS

We would like to gratefully acknowledge the assistance of Heike Jagode and Anthony Danalis (Innovative Computing Lab, UT Knoxville), and Pavan Balaji (Argonne National Lab) in preparing this work. Supported by NSF CCF 1841552.

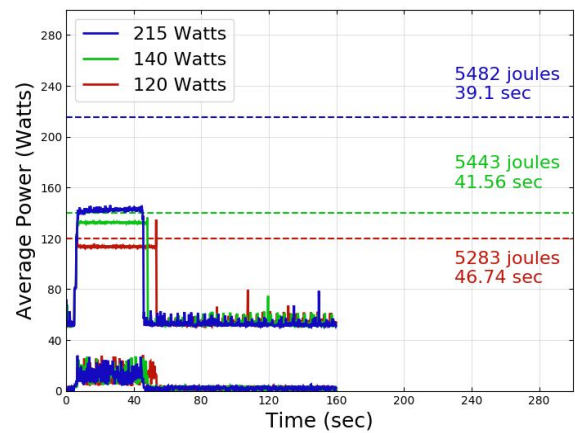

(a)

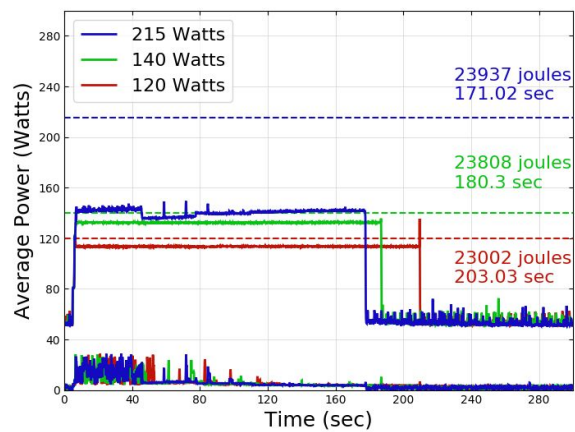

(b)

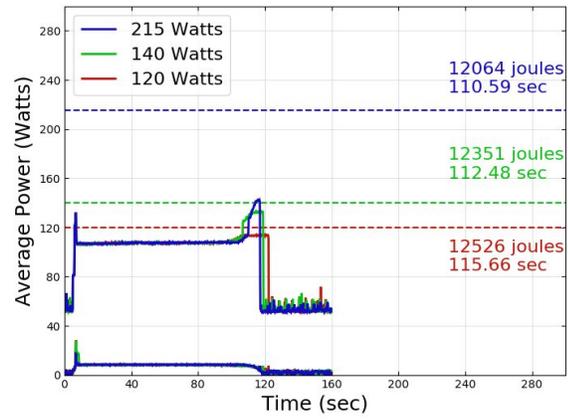

(c)

Figure 1: Performance metrics for the three mini-apps, run on KNM with 72 unique words